\documentclass[aps,pre,twocolumn]{revtex4}

\ifx\pdftexversion\undefined
  \usepackage[dvips]{graphics}
  \DeclareGraphicsExtensions{.eps}
\else
  \usepackage[pdftex]{graphics}
  \DeclareGraphicsExtensions{.pdf}
\fi

\usepackage{amsmath}
\usepackage{amsfonts}
\usepackage{amssymb}

\begin{document}

\title{Dephasing representation of quantum fidelity for general pure and
mixed states}
\author{Ji\v{r}\'{\i} Van\'{\i}\v{c}ek}
\email{vanicek@post.harvard.edu}
\affiliation{Department of Chemistry and Kenneth S. Pitzer Center for Theoretical
Chemistry, University of California, Berkeley, CA 94720}
\date{\today}

\begin{abstract}
General semiclassical expression for quantum fidelity (Loschmidt echo) of arbitrary pure and mixed states is derived. It expresses fidelity as an
interference sum of dephasing trajectories weighed by the Wigner function of
the initial state, and does not require that the initial state be localized
in position or momentum. This general \emph{dephasing representation} is
special in that, counterintuitively, all of fidelity decay is due to
dephasing and none due to the decay of classical overlaps. Surprising accuracy of the approximation is justified by invoking the shadowing theorem: \emph{twice}--both for physical perturbations and for numerical errors. It is shown how
the general expression reduces to the special forms for position and
momentum states and for wave packets localized in position or momentum. The
superiority of the general over the specialized forms is explained and
supported by numerical tests for wave packets, non-local pure states, and for simple and random mixed states. The tests are done in non-universal regimes in mixed phase space where detailed features of fidelity are important. Although semiclassically motivated, present approach is valid for abstract systems with a finite Hilbert basis provided that the discrete Wigner transform is used. This makes the method applicable, via a phase space approach, e. g., to problems of quantum computation.\end{abstract}

\maketitle

\section{\label{sec:intro}Introduction}

Time evolution in classical mechanics is very sensitive to perturbations of
both initial conditions of a trajectory and the Hamiltonian. Because of the
unitarity of quantum evolution, on the other hand, the overlap of two
different quantum states remains constant in time. However, we can still
define sensitivity of quantum evolution to perturbations of the Hamiltonian.
This is usually done using the notion of \emph{quantum fidelity} (sometimes called \emph{Loschmidt echo}), defined for pure
states as \cite{peres:1984}%
\begin{equation}
M\left( t\right) =\left\vert \left\langle \psi\left\vert e^{+iH^{\epsilon
}t/\hbar}e^{-iH^{0}t/\hbar}\right\vert \psi\right\rangle \right\vert ^{2}. 
\label{fidelity}
\end{equation}
Here $|\psi\rangle$ is the initial state, $H^{0}$ and $H^{\epsilon}=H^{0}+%
\epsilon V$ are the unperturbed and perturbed Hamiltonians, respectively. In
words, fidelity is the overlap at time $t$ of two identical initial states
evolved with two slightly different Hamiltonians. Because of its relevance
in theories of decoherence and in experimental realizations of quantum
computation \cite{nielsen:2000}, quantum \cite{jalabert:2001,jacquod:2001,cerruti:2002,prosen:2002,prosen:2002a,jacquod:2003,silvestrov:2003,prosen:2003,prosen:2005,bevilaqua:2004,vanicek:2003a, vanicek:2004a,vanicek:2004b,cucchietti:2002,cucchietti:2002a,wisniacki:2002a, wisniacki:2002,wang:2002,prosen:2002b,jacquod:2002,emerson:2002,wisniacki:2003,cerruti:2003, cucchietti:2003,adamov:2003,kottos:2003,znidaric:2003,hiller:2004,gorin:2004, jacquod:2004,iomin:2004,wang:2004,wang:2005,combescure:2005,weinstein:2005}
as well as classical \cite{benenti:2003,eckhardt:2003,benenti:2003a,garcia:2003,veble:2004}
fidelity has been extensively studied in the last few years. Many universal
regimes of fidelity decay have been found in different limiting cases \cite%
{jalabert:2001,jacquod:2001,cerruti:2002,prosen:2002,prosen:2002a,jacquod:2003, silvestrov:2003,prosen:2003,prosen:2005,bevilaqua:2004}%
. Many of these works used a semiclassical approach, but before Ref. \cite%
{vanicek:2003a} only as a starting point for various approximations, because
of difficulties in treating an exponentially growing number of terms in the
general semiclassical expression for fidelity, especially in chaotic
systems. This problem was solved in Ref. \cite{vanicek:2003a} by a uniform
expression for fidelity which implicitly summed over all these contributions
using an integral over initial conditions, similar in spirit to Miller's
initial value representation \cite{miller:1970,miller:2001}. This
surprisingly simple and accurate expression, although limited to wave
packets localized in position, has been successfully applied as a starting
point to derive fidelity decay in the deep Lyapunov regime \cite{wang:2005}
and the plateau of fidelity in neutron scattering \cite{bevilaqua:2004}.
Five other known regimes of fidelity can also be simply described by this
method \cite{vanicek:2004b}.

In a recent Rapid Communication \cite{vanicek:2004a}, the uniform expression
for fidelity was justified by the shadowing theorem of classical mechanics 
\cite{hammel:1987,grebogi:1990} and a more general and, in fact, \emph{always} more
accurate expression, valid for arbitrary pure states, was stated. One
purpose of the present article is to provide (in Sec. \ref{sec:derivation})
a detailed derivation of this general semiclassical expression (\ref%
{DR_final}) for fidelity of arbitrary pure, i. e., also nonlocal states.
Fidelity is expressed as an interference sum of dephasing trajectories
weighed by the Wigner function of the initial state. The general derivation
provides an alternative and more explicit justification of the validity of
this \emph{dephasing representation} (DR). Interestingly, in Sec. \ref{sec:mixed}
it is shown that the same dephasing representation is valid also for general
mixed states. Section \ref{sec:special} shows how the general expression
reduces to the original form \cite{vanicek:2003a} and other specialized
forms for position and momentum states or Gaussian wave packets localized in
position or momentum. In Sec. \ref{sec:examples}, the general dephasing
representation is tested on a non-local state--a coherent superposition of
two separated wave packets--and on two two types of mixed state--an
incoherent superposition of two wave packets and a completely random state.
It is also shown that the general expression is superior to the original
form \cite{vanicek:2003a} even for a single Gaussian wave packet. All
numerical calculations are done for a system with a finite Hilbert basis. In
such systems, quantum phase space can be rigorously defined if the original
Wigner function \cite{wigner:1932} is replaced by the discrete Wigner
transform \cite{wooters:1987,leonhardt:1995,hannay:1980,rivas:1999,bouzouina:1996}. Since
this discrete transform can be defined in a general abstract Hilbert space
with finite basis, the present approach should be applicable to problems of
quantum computation if phase space approach is used \cite{miquel:2002}. In Sec. \ref{sec:comparison}, DR is compared to other ``Wigner'' methods. The
main conclusions of the paper are summarized in Sec. \ref{sec:conclusion}.

\section{\label{sec:derivation}Dephasing representation for a general pure
state}

Fidelity amplitude for a general pure state $|\psi \rangle $ can be written
as%
\begin{equation}
O\left( t\right) =\left\langle \psi \left\vert e^{+iH^{\epsilon }t/\hbar
}e^{-iH^{0}t/\hbar }\right\vert \psi \right\rangle .
\label{fidelity_amplitude}
\end{equation}%
In order to derive the general dephasing representation of fidelity, we
could start by replacing the two quantum propagators in Eq. (\ref%
{fidelity_amplitude}) by the corresponding semiclassical Van Vleck
propagators \cite{vanvleck:1928}, as in Refs. \cite%
{vanicek:2003a,vanicek:2004a}. However, we will save some effort if we start
directly from the semiclassical initial value representation (IVR) \cite%
{miller:1970,miller:2001} for the two Van Vleck propagators,%
\begin{align}
e^{-iH^{0}t/\hbar }& \approx \left( 2\pi i\hbar \right) ^{-d/2}\int d\mathbf{%
r}_{0}^{\prime }\int d\mathbf{p}_{0}^{\prime }\left\vert \partial \mathbf{r}%
_{t}^{\prime }\left( \mathbf{r}_{0}^{\prime },\mathbf{p}_{0}^{\prime
}\right) /\partial \mathbf{p}_{0}^{\prime }\right\vert ^{1/2} \nonumber \\ 
& \times e^{iS^{0}\left( 
\mathbf{r}_{0}^{\prime },\mathbf{p}_{0}^{\prime };t\right) /\hbar }|\mathbf{r%
}_{t}^{\prime }\rangle \langle \mathbf{r}_{0}^{\prime }|,
\label{ivr_propagator} \\
e^{+iH^{\epsilon }t/\hbar }& \approx \left( -2\pi i\hbar \right) ^{-d/2}\int
d\mathbf{r}_{0}^{\prime \prime }\int d\mathbf{p}_{0}^{\prime \prime
}\left\vert \partial \mathbf{r}_{t}^{\prime \prime }\left( \mathbf{r}%
_{0}^{\prime \prime },\mathbf{p}_{0}^{\prime \prime }\right) /\partial 
\mathbf{p}_{0}^{\prime \prime }\right\vert ^{1/2}
\nonumber \\
& \times e^{-iS^{\epsilon }\left( 
\mathbf{r}_{0}^{\prime \prime },\mathbf{p}_{0}^{\prime \prime };t\right)
/\hbar }|\mathbf{r}_{0}^{\prime \prime }\rangle \langle \mathbf{r}%
_{t}^{\prime \prime }|.  \notag
\end{align}%
Here $\mathbf{r}_{0}^{\prime },\mathbf{p}_{0}^{\prime }$ and $\mathbf{r}%
_{0}^{\prime \prime },\mathbf{p}_{0}^{\prime \prime }$ are the initial
conditions of trajectories of $H^{0}$ and of $H^{\epsilon }$, respectively,
and $\mathbf{r}_{t}^{\prime },\mathbf{p}_{t}^{\prime }$ and $\mathbf{r}%
_{t}^{\prime \prime },\mathbf{p}_{t}^{\prime \prime }$ are the corresponding
coordinates and momenta at time $t$. Action $S^{0\prime }$ of a trajectory
of the unperturbed Hamiltonian $H^{0}$, is given by%
\begin{equation}
S^{0\prime }\left( \mathbf{r}_{0}^{\prime },\mathbf{p}_{0}^{\prime
};t\right) =\int_{0}^{t}d\tau \left[ \mathbf{p}_{\tau }^{\prime }\mathbf{%
\cdot \dot{r}}_{\tau }^{\prime }-H^{0}\left( \mathbf{r}_{\tau }^{\prime },%
\mathbf{p}_{\tau }^{\prime };\tau \right) \right] .  \label{action}
\end{equation}%
Similar expression holds for the action $S^{\epsilon \prime \prime }\left( 
\mathbf{r}_{0}^{\prime \prime },\mathbf{p}_{0}^{\prime \prime };t\right) $
of a trajectory of the perturbed Hamiltonian $H^{\epsilon }$. In the
simplified notation above, the square roots of the determinants in Eq. (\ref%
{ivr_propagator}) also include the appropriate Maslov indices \cite{gutzwiller:1990}. Using the
IVR\ expressions (\ref{ivr_propagator}), fidelity amplitude (\ref%
{fidelity_amplitude}) becomes%
\begin{eqnarray}
O_{\text{IVR}}\left( t\right) &=& \left( 2\pi \hbar \right) ^{-d}\int d\mathbf{r%
}_{0}^{\prime }\int d\mathbf{p}_{0}^{\prime }\int d\mathbf{r}_{0}^{\prime
\prime }\int d\mathbf{p}_{0}^{\prime \prime }\left\vert \frac{\partial 
\mathbf{r}_{t}^{\prime }}{\partial \mathbf{p}_{0}^{\prime }}\right\vert
^{1/2} \nonumber \\
& \times & \left\vert \frac{\partial \mathbf{r}_{t}^{\prime \prime }}{\partial 
\mathbf{p}_{0}^{\prime \prime }}\right\vert ^{1/2}\langle \psi |\mathbf{r}%
_{0}^{\prime \prime }\rangle \langle \mathbf{r}_{t}^{\prime \prime }|\mathbf{%
r}_{t}^{\prime }\rangle \langle \mathbf{r}_{0}^{\prime }|\psi \rangle
\,e^{i\left( S^{0\prime }-S^{\epsilon \prime \prime }\right) /\hbar }.
\label{ivr_amplitude}
\end{eqnarray}

\subsection{Uniform semiclassical expression for fidelity}

If we further expand the $\delta$ function in integral (\ref{ivr_amplitude})
as an integral over a dummy momentum $\mathbf{q}$,%
\begin{equation*}
\langle\mathbf{r}_{t}^{\prime\prime}|\mathbf{r}_{t}^{\prime}\rangle
=\delta\left( \Delta\mathbf{r}_{t}\right) =\left( 2\pi\hbar\right) ^{-d}\int
d\mathbf{q\,}e^{i\mathbf{q\cdot\Delta r}_{t}/\hbar}, 
\end{equation*}
we obtain a ``full'' uniform semiclassical expression for fidelity,%
\begin{eqnarray}
O_{\text{unif}}\left( t\right) & = & \left( 2\pi\hbar\right) ^{-2d}\int d\mathbf{r%
}_{0}^{\prime}\int d\mathbf{p}_{0}^{\prime}\int d\mathbf{r}%
_{0}^{\prime\prime}\int d\mathbf{p}_{0}^{\prime\prime}\int d\mathbf{q} \nonumber \\
& \times & 
\left\vert \frac{\partial\mathbf{r}_{t}^{\prime}}{\partial\mathbf{p}%
_{0}^{\prime}}\right\vert ^{1/2}\left\vert \frac{\partial\mathbf{r}%
_{t}^{\prime\prime}}{\partial\mathbf{p}_{0}^{\prime\prime}}\right\vert
^{1/2}\psi^{\ast}\left( \mathbf{r}_{0}^{\prime\prime}\right) \psi\left( 
\mathbf{r}_{0}^{\prime}\right)  \notag \\
& \times & \exp\left\{ \frac{i}{\hbar}\left[ S^{0\prime}-S^{\epsilon\prime%
\prime}+\mathbf{q\cdot\Delta r}_{t}\right] \right\} .   \label{uniform}
\end{eqnarray}
This integral is, formally, semiclassically \textquotedblleft
exact.\textquotedblright\ In particular, it is free of caustics, unlike, e.
g., the Van Vleck propagator. Because it is expressed only in terms of
initial conditions (and dummy momentum $\mathbf{q}$), it appears to be ready
for numerical evaluations. Unfortunately, this integral is highly
oscillatory, and very difficult to compute, especially in many-dimensional
or chaotic systems. Therefore we will take an alternative route, using a
further approximation, but obtain an integral much easier to tackle
numerically.

\subsection{Dephasing representation}

First, let us make a change of variables $\left\{ \mathbf{r}^{\prime },%
\mathbf{r}^{\prime \prime },\mathbf{p}^{\prime },\mathbf{p}^{\prime \prime
}\right\} \rightarrow \left\{ \mathbf{r},\Delta \mathbf{r},\mathbf{p},\Delta 
\mathbf{p}\right\} $ in integral (\ref{ivr_amplitude}). It should be
emphasized that we do not assume $\Delta \mathbf{r}$ or $\Delta \mathbf{p}$
to be small. New variables (averages and differences) are defined for all
times from $0$ to $t$ as%
\begin{align}
\mathbf{r}& =\frac{1}{2}\left( \mathbf{r}^{\prime }+\mathbf{r}^{\prime
\prime }\right) ,  \label{avg_and_diff} \\
\Delta \mathbf{r}& =\mathbf{r}^{\prime \prime }-\mathbf{r}^{\prime },  \notag
\\
\mathbf{p}& =\frac{1}{2}\left( \mathbf{p}^{\prime }+\mathbf{p}^{\prime
\prime }\right) ,  \notag \\
\Delta \mathbf{p}& =\mathbf{p}^{\prime \prime }-\mathbf{p}^{\prime },  \notag
\end{align}%
The Jacobian of this transformation is unity. If we intend to perform
integrals over $\Delta \mathbf{r}$ and $\Delta \mathbf{p}$ first, we can
consider $\mathbf{r}_{0}$ and $\mathbf{p}_{0}$ as fixed for the moment, and
write
\begin{equation}
\left\vert \frac{\partial\mathbf{r}_{t}^{\prime}}{\partial\mathbf{p}%
_{0}^{\prime}}\right\vert ^{1/2}\left\vert \frac{\partial\mathbf{r}%
_{t}^{\prime\prime}}{\partial\mathbf{p}_{0}^{\prime\prime}}\right\vert
^{1/2}=\left\vert \frac{\partial\left( -\Delta\mathbf{r}_{t}\right) }{%
\partial\left( -\Delta\mathbf{p}_{0}\right) }\right\vert ^{1/2}\left\vert 
\frac{\partial\Delta\mathbf{r}_{t}}{\partial\Delta\mathbf{p}_{0}}\right\vert
^{1/2}=\left\vert \frac{\partial\Delta\mathbf{r}_{t}}{\partial\Delta \mathbf{%
p}_{0}}\right\vert ,   \label{det_transf}
\end{equation}
\begin{eqnarray}
O\left( t\right) &=& \left( 2\pi\hbar\right) ^{-d}\int d\mathbf{r}_{0}\int d%
\mathbf{p}_{0}\int d\Delta\mathbf{r}_{0}\int d\Delta\mathbf{p}_{0}\left\vert 
\frac{\partial\Delta\mathbf{r}_{t}}{\partial\Delta\mathbf{p}_{0}}\right\vert \nonumber \\
& \times & 
\psi^{\ast}\left( \mathbf{r}_{0}^{\prime\prime}\right) \delta\left( \Delta%
\mathbf{r}_{t}\right) \psi\left( \mathbf{r}_{0}^{\prime}\right) \,\exp\left[ 
\frac{i}{\hbar}\left( S^{0\prime}-S^{\epsilon\prime\prime }\right) \right] . 
\label{ivr_amplitude1}
\end{eqnarray}
Next we change variables from $\Delta\mathbf{p}_{0}$ to $\Delta\mathbf{r}_{t}
$ and eliminate the $\delta$ function,%
\begin{eqnarray}
O\left( t\right) &=&\left( 2\pi\hbar\right) ^{-d}\int d\mathbf{r}_{0}\int d%
\mathbf{p}_{0}\int d\Delta\mathbf{r}_{0}\psi^{\ast}\left( \mathbf{r}%
_{0}^{\prime\prime}\right) \psi\left( \mathbf{r}_{0}^{\prime}\right)
\, \nonumber \\
& \times & \left. \exp\left[ \frac{i}{\hbar}\left( S^{0\prime}-S^{\epsilon
\prime\prime}\right) \right] \right\vert _{\Delta\mathbf{r}_{t}=0}. 
\label{ivr_amplitude2}
\end{eqnarray}
The present form is equivalent to Eq. (\ref{uniform}). On one hand, the
present form appears much simpler (a $3d$- vs. $5d$-dimensional integral),
on the other hand it is not an integral over independent variables because
it contains a constraint on the final positions ($\Delta\mathbf{r}_{t}=0$).

While we do not intend to evaluate this integral by the stationary phase
(SP) approximation, it is instructive to check where the action difference $%
S^{0\prime}-S^{\epsilon\prime\prime}$ is stationary because those regions
give the main contributions to the integral. Variation of action $S^{0\prime}
$ gives%
\begin{equation*}
\delta S^{0\prime}=-\mathbf{p}_{0}^{\prime}\cdot\delta\mathbf{r}_{0}^{\prime
}+\mathbf{p}_{t}^{\prime}\cdot\delta\mathbf{r}_{t}^{\prime}
\end{equation*}
and a similar expression holds for $\delta S^{\epsilon\prime\prime}$. Due to the $\Delta\mathbf{r}_{t}=0$ constraint, we have a constraint $\delta\mathbf{r}%
_{t}^{\prime}=\delta\mathbf{r}_{t}^{\prime\prime}$ on the variation of endpoints, and therefore
\begin{equation*}
\delta\left( S^{0\prime}-S^{\epsilon\prime\prime}\right) =-\mathbf{p}%
_{0}^{\prime}\cdot\delta\mathbf{r}_{0}^{\prime}+\mathbf{p}_{0}^{\prime\prime
}\cdot\delta\mathbf{r}_{0}^{\prime\prime}-\Delta\mathbf{p}_{t}\cdot \delta%
\mathbf{r}_{t}^{\prime}. 
\end{equation*}

Expanding variation $\delta\mathbf{r}_{t}^{\prime}$ in terms of variations $\delta 
\mathbf{r}_{0}^{\prime}$ and $\delta\mathbf{p}_{0}^{\prime}$, we find%
\begin{eqnarray}
\delta\left( S^{0\prime}-S^{\epsilon\prime\prime}\right) &=&\left( \Delta%
\mathbf{p}_{0}-\Delta\mathbf{p}_{t}\cdot\frac{\partial\mathbf{r}_{t}^{\prime}%
}{\partial\mathbf{r}_{0}^{\prime}}\right) \cdot\delta \mathbf{r}%
_{0}^{\prime} \notag \\
&-& \Delta\mathbf{p}_{t}\cdot\frac{\partial\mathbf{r}_{t}^{\prime}%
}{\partial\mathbf{r}_{0}^{\prime}}\cdot\delta\mathbf{p}_{0}^{\prime} 
+ \mathbf{%
p}_{0}^{\prime\prime}\cdot\delta\Delta\mathbf{r}_{0}. 
\label{variation_action_diff}
\end{eqnarray}
Note again that so far we have not assumed anything about closeness of the two
trajectories. Since we can easily shift integration variables $\mathbf{r}_{0}
$ and $\mathbf{p}_{0}$ to $\mathbf{r}_{0}^{\prime}$ and $\mathbf{p}%
_{0}^{\prime }$ in Eq. (\ref{ivr_amplitude2}), variation (\ref%
{variation_action_diff}) indeed tells us where the action difference would
be stationary. There are three stationary phase conditions,%
\begin{align}
\Delta\mathbf{p}_{0}-\Delta\mathbf{p}_{t}\cdot\frac{\partial\mathbf{r}%
_{t}^{\prime}}{\partial\mathbf{r}_{0}^{\prime}} & =0, \label{sp_conds}\\
\Delta\mathbf{p}_{t}\cdot\frac{\partial\mathbf{r}_{t}^{\prime}}{\partial 
\mathbf{r}_{0}^{\prime}} & =0, \\
\mathbf{p}_{0}^{\prime\prime}\cdot\delta\Delta\mathbf{r}_{0} & =0.
\end{align}
The third SP condition was intentionally written in the full form. In
general, all three conditions would be satisfied only for a discrete set of
trajectories ($3d$ equations for $3d$ unknowns). However, if the
perturbation were $\epsilon=0$, one could immediately guess that there is
one continuous set of solutions satisfying $\Delta\mathbf{p}_{0}=\Delta%
\mathbf{p}_{t}=\Delta\mathbf{r}_{0}$. The first two conditions are satisfied exactly, the third one approximately for small variations $\delta\Delta\mathbf{r}_{0}$. Even though the third condition is satisfied only approximately, we obtain the correct result--identical trajectories $\Delta\mathbf{r}_{\tau}=0$ for all times $\tau$, $0<\tau<t$--and as we shall see below, also the final result for fidelity will become exact in this limit ($\epsilon = 0$). 
If we add the perturbation, these precise solutions break down, due to the
exponential sensitivity of classical dynamics. However, as was shown in Ref. 
\cite{vanicek:2004a}, if the shadowing theorem \cite%
{hammel:1987,grebogi:1990} is applicable in the given system (for
a given perturbation $\epsilon$ and up to time $t$), there will be a very near
solution with $\Delta\mathbf{r}_{\tau}\approx0$ for all times $\tau$, $%
0<\tau<t$. Putting off a discussion of the shadowing theorem until later,
suffice it to say that this theorem, completely counterintuitively,
guarantees that we can compensate one exponential sensitivity (to
perturbations of $H^{0}$) by another exponential sensitivity (to initial
conditions) and get a trajectory which remains very close to the unperturbed
trajectory up to time $t$. In fact, these approximate (\textquotedblleft
diagonal\textquotedblright) solutions with $\Delta \mathbf{r}_{\tau}\approx0$
will be by far the most dominant ones because for short times no other
solutions exist and for long times the diagonal solutions dephase much
slower than the remaining (\textquotedblleft off-diagonal\textquotedblright)
solutions with different trajectories. Again this will be justified later in
this section. Assuming the validity of shadowing, the \textquotedblleft
diagonal\textquotedblright\ solutions dephase as
\begin{align}
S^{0\prime}-S^{\epsilon\prime\prime} & \approx\epsilon\int_{0}^{t}d\tau
V\left( \mathbf{r}_{\tau},\tau\right) -\Delta\mathbf{r}_{t}\cdot \mathbf{p}%
_{t}+\Delta\mathbf{r}_{0}\cdot\mathbf{p}_{0}  \label{action_diff} \\
& =-\Delta S_{t}-\Delta\mathbf{r}_{t}\cdot\mathbf{p}_{t}+\Delta\mathbf{r}%
_{0}\cdot\mathbf{p}_{0}.
\end{align}
The first term is due to the perturbing potential $\epsilon V$ along the
unperturbed trajectory, the other two terms are due to the small difference
of trajectories at time $t$ and at time $0$. Substituting this action
difference into integral (\ref{ivr_amplitude2}), we obtain the dephasing
representation%
\begin{eqnarray}
O_{\text{DR}}\left( t\right) &=& \left( 2\pi\hbar\right) ^{-d}\int d\mathbf{r}%
_{0}\int d\mathbf{p}_{0}\int d\Delta\mathbf{r}_{0} \notag \\ &\times& \psi^{\ast }\left( \mathbf{%
r}_{0}+\frac{1}{2}\Delta\mathbf{r}_{0}\right) \psi\left( \mathbf{r}_{0}-%
\frac{1}{2}\Delta\mathbf{r}_{0}\right) \notag \\ &\times& 
\exp\left[ \frac {i}{\hbar}\left( -\Delta S_{t}+\Delta\mathbf{r}%
_{0}\cdot\mathbf{p}_{0}\right) \right] . \label{DR_2}
\end{eqnarray}
The final result is more succinctly written as%
\begin{equation}
O_{\text{DR}}\left( t\right) =\int d\mathbf{r}_{0}\int d\mathbf{p}%
_{0}\rho_{W}\left( \mathbf{r}_{0},\mathbf{p}_{0}\right) \,\exp\left(
-i\Delta S_{t}/\hbar\right) ,   \label{DR_final}
\end{equation}
using the Wigner function of the initial state $|\psi\rangle,$ 
\begin{eqnarray}
\rho_{W}\left( \mathbf{r},\mathbf{p}\right) &=& \left( 2\pi\hbar\right)
^{-d}\int d\Delta\mathbf{r\,}\psi^{\ast}\left( \mathbf{r}+\frac{1}{2}\Delta%
\mathbf{r}\right) \notag \\ 
&\times& \psi\left( \mathbf{r}-\frac{1}{2}\Delta \mathbf{r}\right) 
\,\exp\left( i\Delta\mathbf{r}\cdot\mathbf{p}/\hbar\right) . 
\label{wigner_f}
\end{eqnarray}

The general expression (\ref{DR_final}) expresses fidelity as an
interference integral over initial positions $\mathbf{r}_{0}$ and momenta $%
\mathbf{p}_{0}$. Because of this property it was called dephasing
representation in Ref.\cite{vanicek:2004a}. The amplitude of each term is
given by the Wigner function $\rho_{W}\left( \mathbf{r}_{0},\mathbf{p}%
_{0}\right) $ and the phase by the integral of the perturbing potential
along the unperturbed trajectory, $\Delta S_{t}\left( \mathbf{r}_{0},\mathbf{%
p}_{0}\right) $. This is a very intuitive and simple picture that differs
from the simplest \textquotedblleft semiclassical\textquotedblright\ picture
only in using the Wigner function instead of the classical phase space
distribution $\rho_{\text{class}}\left( \mathbf{r}_{0},\mathbf{p}_{0}\right) 
$.

For zero perturbations, $\epsilon=0$, expression (\ref{DR_final})
correctly reduces to the obvious exact result,%
\begin{equation}
O_{\text{DR}}^{\epsilon=0}\left( t\right) =\int d\mathbf{r}_{0}\int d\mathbf{%
p}_{0}\rho_{W}\left( \mathbf{r}_{0},\mathbf{p}_{0}\right) =1 \label{no_pert}
\end{equation}
for all times $t$, where the basic property of the Wigner function was used.

Although we started our derivation for a pure state, we ended up with a
dephasing representation in terms of the Wigner function. Since this
function can also be defined for mixed states, it appears that expression (%
\ref{DR_final}) should remain valid for mixed states, with appropriate
generalization of the notion of fidelity. In Section \ref{sec:mixed}, it
will be shown that this is indeed the case.

\subsection{Shadowing theorem and its double use}

\subsubsection{Trajectories of $H^0$ and $H^{\epsilon}$}

Shadowing theorems in general state that (under certain detailed conditions)
for small enough $\epsilon$ there is a time $t$ such that for a trajectory
of $H^{0}$ with initial condition $\mathbf{r}_{0}^{\prime}$, $\mathbf{p}%
_{0}^{\prime}$ there exists a trajectory of $H^{\epsilon}$ with initial
condition $\mathbf{r}_{0}^{\prime\prime}$, $\mathbf{p}_{0}^{\prime\prime}$
remaining within a certain small distance from the first trajectory up to
time $t$. In uniformly hyperbolic systems this shadowing time $t$ is
infinite \cite{anosov:1967,bowen:1975}, in more general systems at least
finite \cite{grebogi:1990}. Since it is very difficult to find the maximum
shadowing time $t$ and the corresponding bound on the closeness of
trajectories for a specific system, the derivation of dephasing
representation of fidelity assumed that shadowing was applicable for a given
perturbation and time: the numerical results will provide the final
verification.

\subsubsection{Numerical evaluation}

In order to use DR in numerical applications, one only needs to generate
initial conditions $\mathbf{r}_{0}$, $\mathbf{p}_{0}$ from a distribution
given by the Wigner function $\rho_{W}$, run trajectories with the
unperturbed Hamiltonian $H^{0}$ and compute the action difference $\Delta
S_{t}=-\epsilon\int_{0}^{t}d\tau V\left( \mathbf{r}_{\tau},\tau\right) $
along this trajectory. There is no need to compute Van Vleck determinants or
Maslov indices as in many other semiclassical applications. Because the
Wigner function, unlike classical probability, can be negative, some care
must be taken to sample from its distribution. The simplest possible recipe
would be to sample according to the probability $\left\vert
\rho_{W}\right\vert $ and attach a sign afterward together with the
dephasing factor. As we will see from the analysis of special cases in Sec. %
\ref{sec:special}, Wigner function is particularly simple for position and
momentum eigenstates (just a delta function), for Gaussian wave packets (a
Gaussian in both position and momentum), or for a random mixed state (a
constant over the whole phase space). These distributions can be easily
sampled using standard methods. For general pure or mixed states one can
resort to a Monte-Carlo procedure, e. g., using the Metropolis algorithm,
which is frequently done for the IVR approximation \cite{miller:2001}.

One might object that numerical computation of trajectories, due to the exponential sensitivity of classical evolution, will destroy the validity of the DR (\ref{DR_final}). However, here the shadowing theorem helps again--in fact in its original form \cite{hammel:1987,grebogi:1990} where the perturbation was indeed due to errors of numerical propagation. The shadowing idea, as stated in Refs. \cite{hammel:1987,grebogi:1990} guarantees that for each numerical (noisy) trajectory there will be a nearby exact trajectory of $H^{0}$.

\subsection{Comparison of diagonal and off-diagonal terms}

Let us attempt to quantify the validity of the DR by comparing the importance of diagonal and off-diagonal terms in fidelity amplitude. These should be distinguished from the ``diagonal'' and ``off-diagonal'' terms in the fidelity itself (i.e., the amplitude squared), which have been frequently discussed in the literature, see, e. g. Refs. \cite{jalabert:2001,jacquod:2003} where the off-diagonal terms in the fidelity amplitude are already neglected. 

For short enough times $t$, it is clear why DR is accurate--there will be no off-diagonal contributions because there will be no off-diagonal SP solutions of Eqs. (\ref{sp_conds}). For long
times, the number of off-diagonal solutions increases, but in the
semiclassical limit (small $\hbar$) and for small perturbations $\epsilon$,
their contribution is again negligible, due to their much faster dephasing.
Let us see in detail how this happens.

If the unperturbed potential is denoted $W$, then the off-diagonal solutions
dephase as%
\begin{equation}
\Delta S_{\text{off-diag}}=S^{0\prime}-S^{0\prime\prime}=\int_{0}^{t}d\tau%
\left[ W\left( \mathbf{r}_{\tau}^{\prime}\right) -W\left( \mathbf{r}%
_{\tau}^{\prime\prime}\right) \right]    \label{action_diff_offdiagonal}
\end{equation}
because for small enough $\epsilon$, the perturbation $V$ is really
unimportant in dephasing of off-diagonal terms.
Assuming for simplicity that the off-diagonal terms have the same weight and
that their action differences are Gaussian distributed, their average will be%
\begin{equation}
\left\langle e^{i\Delta S_{\text{off-diag}}/\hbar}\right\rangle \approx \exp%
\left[ -\left\langle \left( \Delta S_{\text{off-diag}}\right)
^{2}\right\rangle /2\hbar^{2}\right] .   \label{average_offdiagonal}
\end{equation}
In chaotic systems, $\left\langle \left( \Delta S_{\text{off-diag}}\right)
^{2}\right\rangle =2K_{W}t$, where the diffusion coefficient is $%
K_{W}=\int_{0}^{\infty}dt\,C_{W}\left( t\right) $ and $C_{W}$ is the
potential correlator, $C_{W}\left( t\right) =\left\langle W\left[ \mathbf{r}%
\left( t\right) \right] W\left[ \mathbf{r}\left( 0\right) \right]
\right\rangle $.

Similar analysis can be done for the diagonal terms \cite%
{jalabert:2001,cerruti:2002,vanicek:2003a}. Their average is then given by a
formula analogous to Eq. (\ref{average_offdiagonal}), except with $\Delta S_{%
\text{diag}}$ given by expression (\ref{action_diff}). The variance is now
given by $\left\langle \left( \Delta S_{\text{diag}}\right)
^{2}\right\rangle =2K_{V}\epsilon^{2}t$. Because the diagonal contributions
are weighed by the Wigner function, their total contribution is roughly
equal to the average. The number of discrete off-diagonal semiclassical
contributions should for long times grow as $e^{\gamma t}$ where $\gamma$ is
the topological entropy. Then in the worst possible scenario, where each
off-diagonal term contributes by its full weight (as if the Wigner
function--in the case of diagonal terms--were unity everywhere), the ratio
of the sum of the off-diagonal contributions to the total contribution of
the diagonal terms should be%
\begin{equation*}
\frac{\text{off-diag.}}{\text{diag.}}\sim\exp\left\{ \left[ -\left(
K_{W}-K_{V}\epsilon^{2}\right) /\hbar^{2}+\gamma\right] t\right\} . 
\end{equation*}
For small enough $\epsilon$ and small enough $\hbar$, the off-diagonal terms
will become negligible. Namely, the diagonal terms will give a smaller
contribution if both $\hbar^{2}<K_{W}/\gamma$ and $\epsilon^{2}<\left(
K_{W}-\hbar^{2}\gamma\right) /K_{V}$.

Similar analysis is possible for integrable systems \cite%
{prosen:2002a,vanicek:2004b}. There the number of off-diagonal contributions
grows only algebraically, $\sim t^{\alpha}$ and variance of their action
difference $\left\langle \left( \Delta S_{\text{off-diag}}\right)
^{2}\right\rangle =C_{W}^{\infty}t^{2}$ where $C_{W}^{\infty}=\lim_{t%
\rightarrow\infty}t^{-1}\int_{0}^{t}d\tau\,C_{W}\left( \tau\right) $.
Similarly, for diagonal terms, $\left\langle \left( \Delta S_{\text{diag}%
}\right) ^{2}\right\rangle =C_{V}^{\infty}t^{2}\epsilon^{2}$ \cite{prosen:2002a}. In this case,
the ratio of the two types of contributions is%
\begin{equation*}
\frac{\text{off-diag.}}{\text{diag.}}\sim t^{\alpha}\exp\left[ -\left(
C_{W}^{\infty}-C_{V}^{\infty}\epsilon^{2}\right) t/\left( 2\hbar^{2}\right) %
\right] 
\end{equation*}
and the condition for negligibility of the off-diagonal terms in the limit $%
t\rightarrow\infty$ is $\epsilon^{2}<C_{W}^{\infty}/C_{V}^{\infty}$.

\section{\label{sec:mixed}Dephasing representation for a general mixed state}

There are several ways to generalize the pure-state definition (\ref%
{fidelity_amplitude}) of fidelity to mixed states. The simplest
generalization is%
\begin{equation}
O\left( t\right) =\operatorname*{tr}\left( e^{-iH^{0}t/\hbar}\rho
e^{+iH^{\epsilon}t/\hbar}\right)   \label{fid_ampl_mixed}
\end{equation}
where $\rho$ is the density matrix of the mixed state, normalized such that $%
\operatorname*{tr}\rho=1$ \cite{prosen:2002a}. For pure states $\rho
=|\psi\rangle\langle\psi|$, this general definition reduces to the
pure-state definition (\ref{fidelity_amplitude}). One interpretation of the
general expression (\ref{fid_ampl_mixed}) is that the ket vectors evolve
with the unperturbed Hamiltonian $H^{0}$ and the bra vectors with the
perturbed Hamiltonian $H^{\epsilon}$. Another interpretation is that
expression (\ref{fid_ampl_mixed}) is simply an average of fidelity
amplitudes of pure-state components of the given mixed state. This should be
distinguished from the often studied averaged fidelity.

The second possible generalization of the notion of fidelity to mixed states
replaces the expression for fidelity (\ref{fidelity}), rather than fidelity
amplitude (\ref{fidelity_amplitude}) by an expression%
\begin{equation}
M\left( t\right) =\operatorname*{tr}\left[ \rho ^{0}\left( t\right) \rho
^{\epsilon }\left( t\right) \right] =\operatorname*{tr}\left[ \rho (0)\rho (t)%
\right] ,  \label{fid_mixed2}
\end{equation}%
where $\rho ^{0}\left( t\right) $, $\rho ^{\epsilon }\left( t\right) $ are
the evolved density operators,%
\begin{equation*}
\rho ^{\epsilon }\left( t\right) =e^{-iH^{\epsilon }t/\hbar }\rho
e^{+iH^{\epsilon }t/\hbar },
\end{equation*}%
or, alternatively, $\rho (t)$ is the evolved operator 
\begin{equation*}
\rho \left( t\right) =e^{+iH^{\epsilon }t/\hbar
}e^{-iH^{0}t/\hbar }\rho e^{+iH^{0}t/\hbar }e^{-iH^{\epsilon }t/\hbar }.
\end{equation*}%
Again for pure states $\rho =|\psi \rangle \langle \psi |$, definition (\ref{fid_mixed2}) reduces to the pure-state definition (\ref{fidelity}).

Finally there is another, more intuitive but also more complicated
generalization, which uses the notion of \textquotedblleft purity
fidelity\textquotedblright--the trace of the squared reduced density matrix 
\cite{prosen:2002b},%
\begin{equation}
P_{F}\left( t\right) =\operatorname*{tr}_{S}\left[ \operatorname*{tr}_{E}\rho\left(
t\right) \right] ^{2},   \label{purity_fidelity}
\end{equation}
where subscripts $E$ or $S$ denote that the trace operation is performed on
the environment or system degrees of freedom, respectively. For details see
Ref. \cite{prosen:2002b}. Purity fidelity (\ref{purity_fidelity}) does not, of course, reduce to the definition of fidelity for pure states (\ref{fidelity}).

While dephasing representation expressions are possible for the last two
generalizations, in what follows the simplest generalization (\ref%
{fid_ampl_mixed}) is assumed. With the mixed-state definition (\ref%
{fid_ampl_mixed}), the semiclassical derivation in Eqs. (\ref{ivr_propagator}%
)-(\ref{DR_final}), can be followed closely for mixed states, if we replace
the product $\langle\psi|\mathbf{r}_{0}^{\prime\prime }\rangle\langle\mathbf{%
r}_{0}^{\prime}|\psi\rangle=\psi^{\ast}\left( \mathbf{r}_{0}^{\prime\prime}%
\right) \psi\left( \mathbf{r}_{0}^{\prime }\right) $ in Eqs. (\ref%
{ivr_amplitude}), (\ref{ivr_amplitude1}), and (\ref{DR_2}) by the matrix
element $\langle\mathbf{r}_{0}^{\prime}|\rho|\mathbf{r}_{0}^{\prime\prime}%
\rangle$ of the density operator. For instance, Eq. (\ref{ivr_amplitude1})
will become%
\begin{eqnarray*}
O\left( t\right) &=& \left( 2\pi\hbar\right) ^{-d}\int d\mathbf{r}_{0}\int d%
\mathbf{p}_{0}\int d\Delta\mathbf{r}_{0}\int d\Delta\mathbf{p}_{0}\left\vert 
\frac{\partial\Delta\mathbf{r}_{t}}{\partial\Delta\mathbf{p}_{0}}\right\vert
\, \\ & & \times\,\delta\left( \Delta\mathbf{r}_{t}\right) \,\langle\mathbf{r}_{0}^{\prime
}|\rho|\mathbf{r}_{0}^{\prime\prime}\rangle\,\exp\left[ \frac{i}{\hbar }%
\left( S^{0}-S^{\epsilon}\right) \right] . 
\end{eqnarray*}
At the end, we obtain the same final result (\ref{DR_final}), only the
Wigner function of a pure state (\ref{wigner_f}) must be replaced by the
Wigner-Weyl transform of the density operator,%
\begin{eqnarray}
\rho_{W}\left( \mathbf{r},\mathbf{p}\right) &=&\left( 2\pi\hbar\right)
^{-d}\int d\Delta\mathbf{r\,}\left\langle \mathbf{r}+\frac{1}{2}\Delta%
\mathbf{r}\left\vert \rho\right\vert \mathbf{r}-\frac{1}{2}\Delta\mathbf{r}%
\right\rangle \notag \\
& & \times \exp\left( i\Delta\mathbf{r}\cdot \mathbf{p}/\hbar\right) . 
\label{wigner_f_rho}
\end{eqnarray}

\section{\label{sec:special}Special cases}

For a \emph{position state} $|\mathbf{R}\rangle$, $\psi\left( \mathbf{r}%
\right) =\delta\left( \mathbf{r}-\mathbf{R}\right) $, the Wigner function (%
\ref{wigner_f}) is 
\begin{align}
\rho_{W}^{\text{pos.st.}}\left( \mathbf{r},\mathbf{p}\right) & =\left(
2\pi\hbar\right) ^{-d}\int d\mathbf{x\,}\delta\left( \mathbf{r}+\frac{1}{2}%
\mathbf{x-R}\right) \notag \\ 
&\ \times \delta\left( \mathbf{r}-\frac{1}{2}\mathbf{x-R}\right)
\,\exp\left( i\mathbf{x}\cdot\mathbf{p}/\hbar\right)  \notag
\\
& =\left( 2\pi\hbar\right) ^{-d}\mathbf{\,}\delta\left( \mathbf{r-R}\right) .
\label{wig_pos_state}
\end{align}
Substituting Eq. (\ref{wigner_f}) into the general dephasing representation (%
\ref{DR_final}), we find%
\begin{equation*}
O_{\text{DR}}^{\text{pos.st.}}\left( t\right) =\left( 2\pi\hbar\right)
^{-d}\int d\mathbf{p}_{0}\,\exp\left[ -i\Delta S_{t}\left( \mathbf{R},%
\mathbf{p}_{0}\right) /\hbar\right] , 
\end{equation*}
in agreement with Eq. (1) from Ref. \cite{vanicek:2004a} and with \cite%
{vanicek:2003a,wang:2004}.

For a \emph{momentum state} $|\mathbf{P\rangle}$, $\psi\left( \mathbf{r}%
\right) =\left( 2\pi\hbar\right) ^{-d/2}\exp\left( i\mathbf{P}\cdot\mathbf{%
r/\hbar}\right) $, the Wigner function (\ref{wigner_f}) becomes%
\begin{eqnarray}
\rho_{W}^{\text{mom.st.}}\left( \mathbf{r},\mathbf{p}\right) &=& \left(
2\pi\hbar\right) ^{-2d}\int d\mathbf{x}\exp\left[ i\left( \mathbf{p-P}%
\right) \cdot\mathbf{x/\hbar}\right] \notag \\ &=& \left( 2\pi\hbar\right)
^{-d}\delta\left( \mathbf{p-P}\right) \text{,}   \label{wig_mom_state}
\end{eqnarray}
and the general DR of fidelity (\ref{DR_final}) reduces to%
\begin{equation*}
O_{\text{DR}}^{\text{mom.st.}}\left( t\right) =\left( 2\pi\hbar\right)
^{-d}\int d\mathbf{r}_{0}\,\exp\left[ -i\Delta S_{t}\left( \mathbf{r}_{0},%
\mathbf{P}\right) /\hbar\right] . 
\end{equation*}

A \emph{general Gaussian wave packet} with average position $\mathbf{R}$,
average momentum $\mathbf{P}$, and position spread $\sigma$,%
\begin{equation*}
\psi\left( \mathbf{r}\right) =\left( \pi\sigma^{2}\right) ^{-d/4}\exp\left[ i%
\mathbf{P\cdot}\left( \mathbf{r-R}\right) /\hbar-\left( \mathbf{r-R}\right)
^{2}/2\sigma^{2}\right] 
\end{equation*}
has Wigner function%
\begin{align}
&\rho_{W}^{\text{gen.G.w.p.}}\left( \mathbf{r,p}\right)  =\left(
\pi\sigma^{2}\right) ^{-d/2}\left( 2\pi\hbar\right) ^{-2d}\int d\mathbf{x} \notag \\
& \ \times \exp\left\{ \frac{i}{\hbar}\left( \mathbf{p-P}\right) \cdot\mathbf{x-}\left[
\left( \mathbf{r-R}\right) ^{2}+\left( \mathbf{x}/2\right) ^{2}\right]
/\sigma^{2}\right\}  \notag  \\
& =\left( \pi\hbar\right) ^{-d}\exp\left[ -\left( \mathbf{r-R}\right)
^{2}/\sigma^{2}-\left( \mathbf{p-P}\right) ^{2}\sigma^{2}/\hbar^{2}\right] .
\label{wig_wavepacket}
\end{align}
In general the dephasing representation of a Gaussian wave packet is (\ref%
{DR_final}) with the Wigner function (\ref{wig_wavepacket}) where we must
include dephasing trajectories with varying \emph{both} positions \emph{and }%
momenta. Only in special cases, such as when the wave packet is strongly 
\emph{localized in position} (i. e., when $\sigma\ll\hbar^{1/2}$), can we
make a further simplification by replacement of $\Delta S_{t}\left( \mathbf{r%
}_{0},\mathbf{p}_{0}\right) $ by $\Delta S_{t}\left( \mathbf{R,p}_{0}\right) 
$ in Eq. (\ref{DR_final}). Then we can compute the $\mathbf{r}_{0}$ integral
in Eq. (\ref{DR_final}) analytically and obtain%
\begin{eqnarray}
&& O_{\text{DR}}^{\text{pos.G.w.p.}}\left( t\right) =\left(
\sigma^{2}/\pi\hbar^{2}\right) ^{d/2}\int d\mathbf{p}_{0} \notag \\ &&\ \times \exp\left[ -i\Delta
S_{t}\left( \mathbf{R,p}_{0}\right) /\hbar-\left( \mathbf{p-P}\right)
^{2}\sigma^{2}/\hbar^{2}\right],\ \label{fid_pos_loc}  
\end{eqnarray}
in agreement with Eq. (8) in Ref.\cite{vanicek:2003a}. There the same result
was obtained by linearizing the Van Vleck semiclassical propagator about the
central trajectory. In Section \ref{sec:examples} it will be shown that the
symmetric expression (\ref{wig_wavepacket}) based on the general DR (\ref%
{DR_final}) is superior to the specialized form (\ref{fid_pos_loc}).
Similarly, if the initial Gaussian wave packet is \emph{localized in momentum%
} (i. e., when $\sigma\gg\hbar^{1/2}$), we can replace $\Delta S_{t}\left( 
\mathbf{r}_{0},\mathbf{p}_{0}\right) $ by $\Delta S_{t}\left( \mathbf{r}_{0}%
\mathbf{,P}\right) $ and obtain%
\begin{eqnarray}
&& O_{\text{DR}}^{\text{mom.G.w.p.}}\left( t\right) =\left( \pi\sigma
^{2}\right) ^{-d/2}\int d\mathbf{r}_{0}  \notag \\ && \ \times \exp\left[ -i\Delta S_{t}\left( 
\mathbf{r}_{0}\mathbf{,P}\right) /\hbar-\left( \mathbf{r-R}\right)
^{2}/\sigma^{2}\right].\  \label{fid_mom_loc} 
\end{eqnarray}
For general (non-Gaussian) wave packets, which are nevertheless localized
either in position (about $\mathbf{R}$) or momentum (about $\mathbf{P}$), we
can use the general property of the Wigner function%
\begin{align*}
\int d\mathbf{r\,}\rho_{W}\left( \mathbf{r,p}\right) & =\left\vert
\psi\left( \mathbf{p}\right) \right\vert ^{2}, \\
\int d\mathbf{p\,}\rho_{W}\left( \mathbf{r,p}\right) & =\left\vert
\psi\left( \mathbf{r}\right) \right\vert ^{2},
\end{align*}
and obtain, upon substitution into the general DR (\ref{DR_final}),%
\begin{align}
O_{\text{DR}}^{\text{pos.w.p.}}\left( t\right) & =\int d\mathbf{p}_{0}\exp%
\left[ -i\Delta S_{t}\left( \mathbf{R,p}_{0}\right) /\hbar\right] \left\vert
\psi\left( \mathbf{p}_{0}\right) \right\vert ^{2}, \\
O_{\text{DR}}^{\text{mom.w.p.}}\left( t\right) & =\int d\mathbf{r}_{0}\exp%
\left[ -i\Delta S_{t}\left( \mathbf{r}_{0}\mathbf{,P}\right) /\hbar\right]
\left\vert \psi\left( \mathbf{r}_{0}\right) \right\vert ^{2}.
\end{align}

Finally, for a completely \emph{random state}, i. e., an incoherent
superposition of all pure basis states, the density operator as well as its
Wigner function (\ref{wigner_f_rho}) is just a constant (independent of
position or momenta), and for a system with a finite phase space volume $%
\Omega$, the DR becomes%
\begin{equation}
O_{\text{DR}}^{\text{random st.}}\left( t\right) =\frac{1}{\Omega}\int d%
\mathbf{r}_{0}\int d\mathbf{p}_{0}\exp\left( -i\Delta S_{t}/\hbar\right) . 
\label{fid_random}
\end{equation}

It should be pointed out that while names like \textquotedblleft
position\textquotedblright\ or \textquotedblleft momentum\textquotedblright\
states have been used to describe the special cases, they do not necessarily
need to be eigenstates of the usual position or momentum operator. In the
case of abstract Hilbert space with a finite basis, \textquotedblleft
position\textquotedblright\ states are simply the basis states (called \emph{%
computational} states in the setting of quantum information, could be, e. g.
spin eigenstates), and \textquotedblleft momentum\textquotedblright\ states
are simply the states defined by the discrete Fourier transform of the
original basis states \cite{miquel:2002}. In Ref. \cite{miquel:2002}, this
generalized phase-space representation is used to show that for quite a few
interesting operations on computational states, the Wigner function evolves
classically. In all these cases, the dephasing representation described in
Secs. \ref{sec:derivation}-\ref{sec:examples} should be applicable if
discrete Wigner function \cite{miquel:2002} is used and simple other
modifications are made to account for the finite-size of phase space. In
fact, this is done in the numerical examples in the following section.

\section{\label{sec:examples}Numerical tests}

Now let us apply the theoretical analysis from previous sections to a
specific system, the Chirikov standard map. Its advantage is that it is
discrete, coordinate space is only one-dimensional, but at the same time
standard map already contains generic complexities of classical dynamics.
Specifically, the phase space is mixed and so various simplifications
applicable in quasi-integrable or strongly chaotic systems are in general
not applicable. Standard map is a symplectic map defined on a compact
two-dimensional phase space--torus, as follows, 
\begin{align*}
q_{j+1} & =q_{j}+p_{j}\text{ \ \ (mod }2\pi\text{)} \\
p_{j+1} & =p_{j}-W^{\prime}\left( q_{j+1}\right) -\epsilon V^{\prime }\left(
q_{j+1}\right) \text{ \ \ (mod }2\pi\text{)}\mathrm{,}
\end{align*}
where $q$ and $p$ are position and momentum on the torus, potential $W\left(
q\right) =-k\cos q$, and the perturbation is $V\left( q\right) =-\cos2q$.
Using an $n$-dimensional Hilbert space for the quantized map fixes the
effective Planck constant to be $\hbar=\left( 2\pi n\right) ^{-1}$. (We are
using letter $q$ for the coordinate to distinguish this special system from
the general considerations. Similarly, we will use letter $Q$ to denote the
position of a position state or center of a wave packet.) Parameter $\epsilon
$ controls the strength of perturbation. For $\epsilon\ll1$, the map is
close to being integrable, for $\epsilon\gg1$, the map is strongly chaotic.
The goal of this section is not to use the dephasing representation to
explore various universal regime that occur in these two limits and have
been carefully studied in the literature. This was already done in Refs. 
\cite{vanicek:2003a,vanicek:2004b}. The goal of this section is rather to
explore the detailed features of fidelity in non-universal regimes. The
optimal region of parameter space is in the vicinity of $\epsilon=1$, since
there phase space has a significant amount of chaotic as well as integrable
regions. Mixed phase space is in general the hardest to treat and therefore
this setting is chosen here because it provides the most challenging test
for any approximation.

\subsection{Gaussian wave packets}

One might think that the general dephasing representation (\ref{DR_final})
is only useful for highly non-local states and that the original expression (%
\ref{wig_pos_state}) from Ref. \cite{vanicek:2003a} is good enough at least
for Gaussian wave packets. This subsection demonstrates that even for
Gaussian wave packets, the general dephasing representation (\ref{DR_final})
is superior to the original expression (\ref{wig_pos_state}) from Ref. \cite%
{vanicek:2003a}.

Figure 1 compares three approximations to compute fidelity of Gaussian wave
packets with the exact result: the expression (\ref{wig_pos_state}) from
Ref. \cite{vanicek:2003a} for wave packets localized in position (red dashed
line), corresponding expression (\ref{fid_mom_loc}) for wave packets
localized in momentum (blue dotted line), and the general DR (\ref{DR_final}%
), with the Wigner function (\ref{wig_wavepacket}), symmetrically treating
position and momentum (black solid line). 
\begin{figure}[ptbh]
\centerline{\resizebox{\hsize}{!}{\includegraphics{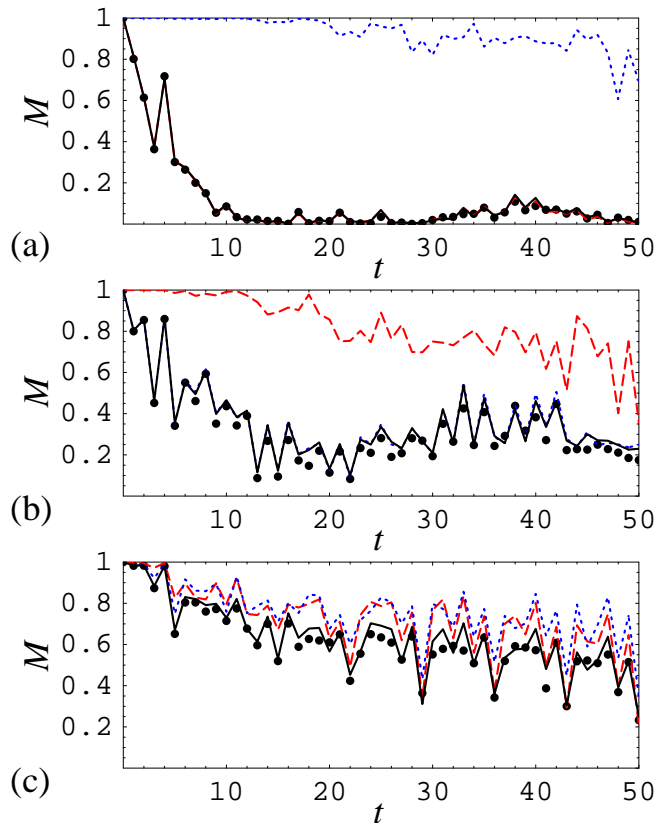}}}
\caption{Fidelity for a Gaussian wave packet centered at $Q = 0.7\protect\pi$%
, $P = 0.4\protect\pi$ in a perturbed standard map with $n=1000$, $k=0.95$, $%
\protect\epsilon=0.015$. Comparison of the exact result (solid dots),
general dephasing representation (black line) and its specialized forms for
position-like (red dashed line) and momentum-like wave packets (blue dotted
line). The initial position spread $\protect\sigma$ of the Gaussian is: a)
0.004$\protect\pi$, b) 0.16$\protect\pi$, and c) 0.04$\protect\pi$. The DR
calculations used $1000$ classical trajectories.}
\end{figure}
The exact fidelity, computed by exact quantum evolution using the Fast
Fourier Transform algorithm, is represented by solid dots. The parameters
are $n=1000$, $k=0.95$, $\epsilon=0.015$, and the wave packet is localized
at $Q=0.7\pi$ and $P=0.4\pi$. The number of classical trajectories used in
the calculations is $1000$. Wave packets used in parts a), b), and c) of
Fig. 1 have position spread $\sigma$ equal to 0.004$\pi$, 0.16$\pi$, and 0.04%
$\pi$, respectively. For a wave packet localized in position in Fig. 1a),
the original expression (\ref{wig_pos_state}) from Ref. \cite{vanicek:2003a}
works very well and is almost indistinguishable from the general DR (\ref%
{DR_final}), as expected, whereas Eq. (\ref{fid_mom_loc}) for momentum wave
packets fails. For a wave packet localized in momentum in Fig. 1b), the
momentum-wave-packet expression (\ref{fid_mom_loc}) works well and it is
almost indistinguishable from the general DR (\ref{DR_final}), but the
original position-wave-packet expression (\ref{wig_pos_state}) from Ref. 
\cite{vanicek:2003a} fails completely. The general DR works very well in
both cases. It might seem that either the momentum or position versions
could cover the whole range of Gaussian wave packets, because one might
think that the intermediate case, i. e., a fairly symmetric wave packet, is
localized enough in both position and momentum. That this is not so is
provided by the final test in Fig. 1c): both specialized expressions (\ref%
{fid_pos_loc}) and (\ref{fid_mom_loc}) give a significant error in
comparison with exact fidelity, but the general DR (\ref{DR_final}) gives
very accurate results, as expected because of its \textquotedblleft
fair\textquotedblright\ treatment of position and momentum. To conclude,
expression (\ref{DR_final}), is accurate for the whole range of Gaussian
wave packets, from position-like to symmetric to momentum-like, even in the
presence of mixed dynamics.

\subsection{Nonlocal states}

For nonlocal states, there is even less hope that the position-wave-packet
expression for fidelity (\ref{fid_pos_loc}) from Ref. \cite{vanicek:2003a}
would work. One might think that for a superposition of localized wave
packets it is enough to simply add the terms (\ref{fid_pos_loc}) for the
fidelity amplitude. This is not the case which can be seen by considering a
wave packet $\psi$ that is a superposition of two Gaussian wave packets $%
\psi_{1}$ and $\psi_{2}$, centered at phase space points $\left( \mathbf{R}%
_{1}\mathbf{,P}_{1}\right) $ and $\left( \mathbf{R}_{2}\mathbf{,P}%
_{2}\right) $. The resulting wave packet has a Wigner function that is not
just a simple sum of the Wigner functions of the two Gaussian wave packets.
The correct Wigner function has in addition an \emph{interference term}
localized in the vicinity of the phase-space point $\left( \mathbf{(R}_{1}+%
\mathbf{R}_{2})/2\mathbf{,(P}_{1}+\mathbf{P}_{2})/2\right) $. We will
demonstrate now the importance of this interference term and show that if it
is taken into account, the general DR (\ref{DR_final}) will still give
excellent results, even for nonlocal states.

Being motivated by the quantum computation applications, let us consider a
superposition of computational states (i. e., position states in the
abstract phase space), instead of Gaussian wave packets. Our initial state
is a \emph{coherent} superposition%
\begin{equation}
|\psi\rangle=\frac{1}{\sqrt{2}}\left( |\mathbf{R}_{1}\rangle+|\mathbf{R}%
_{2}\rangle\right) ,   \label{coh_superpos}
\end{equation}
with a Wigner distribution,%
\begin{align}
&\rho_{W}^{\text{coh}}\left( \mathbf{r,p}\right) =\frac{1}{2}\left(
2\pi\hbar\right) ^{-d} \{ \delta ( \mathbf{r-R}_{1})
+\delta ( \mathbf{r-R}_{2} ) \notag \\
&+ 2\delta\left[ \mathbf{r}-\left( 
\mathbf{R}_{1}+\mathbf{R}_{2}\right) /2\right] \cos\left[ \left( \mathbf{R}%
_{1}-\mathbf{R}_{2}\right) \cdot\mathbf{p/\hbar}\right] \} . 
\label{wig_coher}
\end{align}
If the interference term is neglected, we obtain a Wigner function of the 
\emph{incoherent} superposition (\ref{mixed_state}),%
\begin{equation}
\rho_{W}^{\text{incoh}}\left( \mathbf{r,p}\right) =\frac{1}{2}\left(
2\pi\hbar\right) ^{-d}\left[ \delta\left( \mathbf{r-R}_{1}\right)
+\delta\left( \mathbf{r-R}_{2}\right) \right]   \label{wig_incoh}
\end{equation}

Figure 2 compares two approximate ways to compute fidelity with the exact
quantum result: both approximations use the general DR (\ref{DR_final}), but
whereas one uses the correct full Wigner function (\ref{wig_coher}) (black
solid line), the other uses the incorrect Wigner function (\ref{wig_incoh}),
neglecting the interference term (purple dashed-dotted line). Again, the
exact result is represented by solid dots. The parameters used in Fig. 2 are 
$n=200$, $k=0.7$, $\epsilon=0.02$, $Q_{1}=0.4\pi$, and $400$ classical
trajectories were used. The position of the other component state varies in
the two parts. 
\begin{figure}[ptbh]
\centerline{\resizebox{\hsize}{!}{\includegraphics{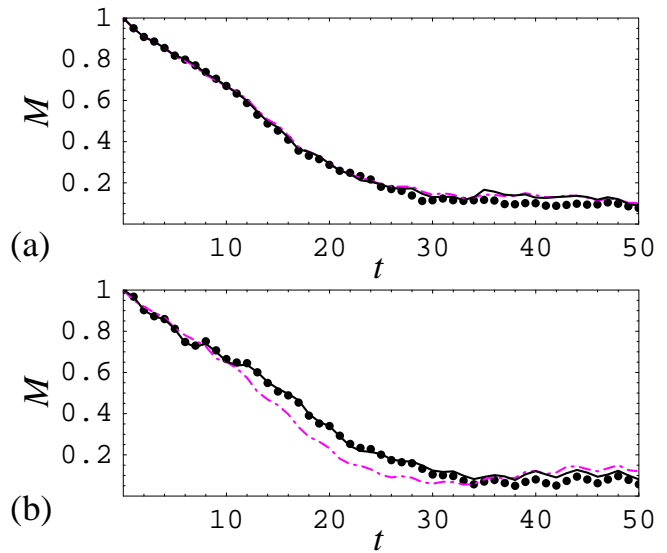}}}
\caption{Fidelity for a nonlocal state--coherent superposition of two
position states--in a perturbed standard map with $n=200$, $k=0.7$, $\protect%
\epsilon=0.02$. Comparison of the exact result (solid dots), general
dephasing representation (black line) and the approximate DR neglecting
coherence effects in the Wigner function (purple dashed-dotted line). The
two position states are located at: a) $Q_{1}=0.4\protect\pi$ and $Q_{2}=1.2%
\protect\pi$, b) $Q_{1}=0.4\protect\pi$ and $Q_{2}=0.42\protect\pi$. In both
parts, $400$ classical trajectories were used.}
\end{figure}

If the positions $\mathbf{R}_{1}$ and $\mathbf{R}_{2}$ are largely
separated, the oscillations in the interference term have a high frequency.
Because nearby initial conditions follow similar trajectories and have
similar actions, the phase factor in the DR (\ref{DR_final}) varies slowly.
Therefore the fast oscillations in the weight factor given by the
interference term in the Wigner function can completely cancel out the
contribution of the interference part to the DR integral. (Incidentally,
this situation is in a way opposite to the usual semiclassical
considerations where the weight is a slowly varying function and the phase
factor is the fast oscillating factor.) Figure 1a) shows an example of
situation where this cancellation occurs: $Q_{1}=0.4\pi$ and $Q_{2}=1.2\pi$.
Because the interference term is negligible, both approximations give the
same and very accurate result.

If the initial states are closer, as in Fig. 2b), where $Q_{1}=0.4\pi$ and $%
Q_{2}=0.42\pi$, the interference term is important, and only the correct
Wigner function (\ref{wig_coher}) agrees well with the exact result. This
shows that for coherent nonlocal states, the general DR (\ref{DR_final})
must be used instead of some approximate versions which neglect quantum
coherence of the initial state.

\subsection{Mixed states}

Wigner function (\ref{wig_incoh}) was wrong for the coherent state (\ref%
{coh_superpos}), but it does correctly describe a certain mixed state,
namely the incoherent superposition of computational states $|\mathbf{R}%
_{1}\rangle$ and $|\mathbf{R}_{2}\rangle$,%
\begin{equation}
\rho^{\text{incoh}}=\frac{1}{2}\left( |\mathbf{R}_{1}\rangle\langle \mathbf{R%
}_{1}|+|\mathbf{R}_{2}\rangle\langle\mathbf{R}_{2}|\right) . 
\label{mixed_state}
\end{equation}
In Sec. \ref{sec:mixed} it was shown that if the generalized definition (\ref%
{fid_ampl_mixed}) of fidelity for mixed states is used, dephasing
representation (\ref{DR_final}) remains valid, as long as the Wigner
transform of the density operator (\ref{wigner_f_rho}) is used. For the
incoherent mixture with density operator (\ref{mixed_state}), Wigner
distribution is precisely that given by Eq. (\ref{wig_incoh}). Figure 3
compares DR (\ref{DR_final}) with the Wigner function (\ref{wig_incoh}) with
the exact fidelity for the state (\ref{mixed_state}). Fidelity computed by
the DR is drawn with a black solid line, exact fidelity with solid dots. The
parameters are the same as in Fig. 2b), in particular $Q_{1}=0.4\pi$ and $%
Q_{2}=0.42\pi$. Although now only $200$ classical trajectories were used,
the agreement is again excellent. 
\begin{figure}[ptbh]
\centerline{\resizebox{\hsize}{!}{\includegraphics{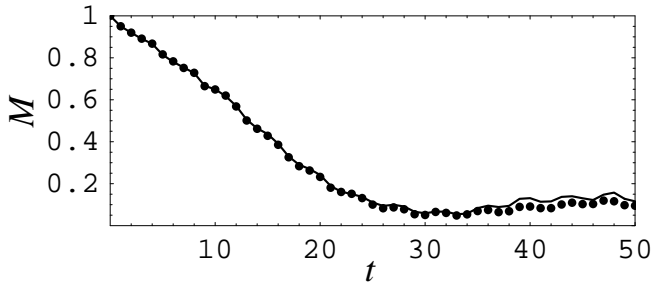}}}
\caption{Fidelity for a mixed state--incoherent superposition of two
position states--in a perturbed standard map. All parameters are the same as
in Fig. 2b), except that only $200$ trajectories were used. Comparison of
the exact result (solid dots) and general dephasing representation (black
line).}
\end{figure}

Last but not least we consider the completely random mixed state. It is an
incoherent superposition of all computational states and in a
finite-dimensional Hilbert space, its density operator is%
\begin{equation*}
\rho^{\text{random}}=\frac{1}{n}\sum_{i=1}^{n}|Q_{i}\rangle\langle Q_{i}|=%
\frac{1}{n}\hat{1}. 
\end{equation*}
Figure 4 compares the random-state version (\ref{fid_random}) of DR (black
solid line) with the exact result (solid dots). Parameters in this
calculation are $n=100$, $k=2$, $\epsilon=0.03$ and $1000$ classical
trajectories were used. Again, it is reassuring that even in the case that
the whole phase space is important, with just $1000$ trajectories, dephasing
representation still works so well--despite the fact that it was derived
solely from semiclassical arguments and requires only classical information. 
\begin{figure}[ptbh]
\centerline{\resizebox{\hsize}{!}{\includegraphics{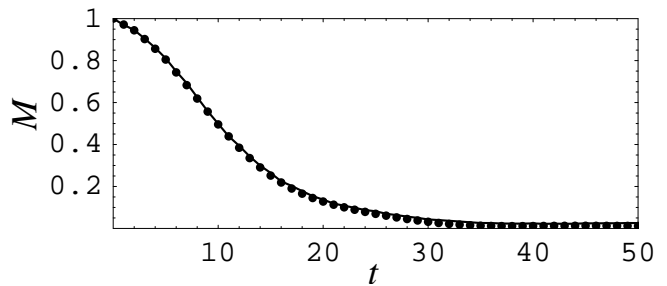}}}
\caption{Fidelity for a random mixed state--incoherent superposition of all
basis states--in a perturbed standard map with $n=100$, $k=2$, $\protect%
\epsilon=0.03$. Comparison of the exact (solid dots) result and general
dephasing representation (black line). $1000$ classical trajectories were
used.}
\end{figure}

\section{\label{sec:comparison}Relation to other ``Wigner'' methods}

It should be noted that the Wigner distribution has been used in various
other approximate methods, especially in chemical physics. For instance, it
was used to compute photodissociation cross-sections 
\cite{heller:1976,brown:1981}, to treat inelastic scattering \cite{lee:1980}%
, or to compute thermal correlation functions using the linearized
semiclassical IVR method \cite{miller:1974a,wang:1998,miller:2001}. In all
these applications, there was just one Hamiltonian, but the two states (or
more generally, density or other operators) were different. The quantity of
interest was a general correlation function of the type
\begin{equation}
C_{AB}\left( t\right) =\operatorname*{tr}\left( AU^{\dag }BU\right) \label{corr_f}
\end{equation}%
where $A$ and $B$ are general operators and $U=\mathcal{T}e^{-i \int H d \tau}$ is the time evolution operator. Using various approximations, all authors \cite{heller:1976,brown:1981,lee:1980,miller:1974a,wang:1998,miller:2001} obtain the same final result, expressed as an overlap of two Wigner distributions, one at
time 0, the other evolved classically to time $t$, 
\begin{equation}
C_{AB}^{\text{Wigner}}\left( t\right) = (2\pi \hbar)^d \int d\mathbf{r}_{0}\int d\mathbf{p}%
_{0}A_{W}\left( \mathbf{r}_{0},\mathbf{p}_{0}\right) B_{W}\left( \mathbf{r}%
_{t},\mathbf{p}_{t}\right) \,
\end{equation}
Here $A_W$ and $A_W$ are the Wigner transforms (\ref{wigner_f_rho}) of operators $A$ and $B$. 

Because there is only one Hamiltonian, there is no dephasing factor $%
e^{i\Delta S/\hbar }$, as in the DR. In fact we could apply one of these older approaches to the second generalized definition (\ref{fid_mixed2}) of fidelity for mixed states because that definition is in the form of Eq. (\ref{corr_f}) with $A=B=\rho$ and the time evolution operator $U = e^{+iH^{\epsilon}t/\hbar}e^{-iH^{0}t/\hbar}$. Then we would obtain a very different result from the DR,
 \begin{equation}
M^{\text{Wigner}}\left( t\right) = (2\pi \hbar)^d \int d\mathbf{r}_{0}\int d\mathbf{p}%
_{0}\rho_{W}\left( \mathbf{r}_{0},\mathbf{p}_{0}\right) \rho_{W}\left( \mathbf{r}%
_{t},\mathbf{p}_{t}\right). \label{fid_wig}
\end{equation}
Although appearing as elegant as the dephasing representation, there is a problem with this expression. First, it will be much more sensitive to numerical errors. We can see that already by considering zero perturbation. Correctly, for each initial condition $\mathbf{r}_{0},\mathbf{p}_{0}$, we should have $\mathbf{r}_{0}=
\mathbf{r}_{t}$ and $\mathbf{p}_{t}=\mathbf{p}_{0}$. In systems with nonlinear dynamics, particularly chaotic systems, numerical errors in forward and backward propagation will yield exponentially growing errors. If the initial state is a localized wave packet, expression (\ref{fid_wig}) would give a numerically decaying overlap even for zero perturbations when exact fidelity is constant $M(t) = 1$. Indeed, numerical test not presented here showed that instead of staying at unity,  $M^{\text{Wigner}}$ quickly decays to a plateau and remains there for some time, and finally decays exponentially again. (This is the same behavior as observed in literature for physical perturbations \cite{prosen:2005,bevilaqua:2004}. 

Even if numerical errors did not exist, equation (\ref{fid_wig}) would have problems. It can describe some decay due to dephasing, but only that in the fast oscillating parts of the initial state. For simple Gaussian wave packets, the fidelity decay in Eq. (\ref{fid_wig}) is completely due to the decay of classical overlaps, i. e., classical fidelity. To conclude, the ``Wigner'' form (\ref{fid_wig}) is apparently not as good as the dephasing representation, but it does deserve further study, especially because it might shed further light on the question of importance of various contributions to fidelity. Preliminary studies show that $M^{\text{Wigner}}$ correctly describes exact fidelity in both chaotic and quasi-integrable systems for large perturbations  (i. e., in Lyapunov and algebraic regimes, respectively), when dephasing is not important \cite{vanicek:2004b}. It gives wrong results in both chaotic and quasi-integrable systems for small perturbations (in the FGR and Gaussian regimes), when dephasing is important \cite{vanicek:2004b}.

\section{\label{sec:conclusion}Conclusion}

This paper has presented a derivation of a general semiclassical expression for fidelity of pure and mixed states. This dephasing representation expresses fidelity as an interference integral, with weight of each term given by the Wigner function and the phase by the integrated perturbation along an unperturbed trajectory.  In particular, no analog of the Van Vleck determinant is needed. As the original specialized expression (\ref{fid_pos_loc}) from Ref. \cite{vanicek:2003a}, dephasing representation avoids searching for the exponentially growing number of terms in the standard semiclassical expressions \cite{jalabert:2001}. It also avoids the ubiquitous divergences in Van Vleck determinants present in the usual semiclassical expressions. 

The advantage of dephasing representation lies in that it does not require the original state to be localized. Its form suggests that it should be applicable to general pure and mixed states. This claim was supported by the following numerical evidence:
First, it was shown, on the example of Gaussian wave packets, that position and momentum must be treated symmetrically. This was the flaw of the expression from Ref. \cite{vanicek:2003a} and is apparently corrected in the DR. Second, on the example of coherent superpositions of states, it was shown that oscillatory patterns in the Wigner function are important: therefore classical phase space distribution, resulting from incoherent superposition of component Wigner distributions (for states for which these are the same as classical distributions).
This may shed some further light on the controversial issue of importance of sub-Plank structures on decoherence \cite{zurek:2001,jacquod:2002}.
Finally, it was shown that DR is also accurate for mixed states: incoherent superpositions and completely random states. All tests were performed on a system with mixed phase space: with both integrable and chaotic regions. 

While the numerical tests were quite successful, a further study is needed to determine precisely all situations where the dephasing representation breaks down. The analysis provided in Sec. \ref{sec:derivation} of this paper should simplify that task. Also, a more rigorous formulation of the precise conditions of validity of the dephasing representation is needed.  

\begin{acknowledgments}
The author wishes to thank the Department of Chemistry at the University of California, Berkeley for support.
\end{acknowledgments}

\end{document}